# Jupiter's Temperature Structure: A Reassessment of the Voyager Radio Occultation Measurements

Pranika Gupta[1], Sushil K. Atreya[1], Paul G. Steffes[2], Leigh N. Fletcher[3], Tristan Guillot[4], Michael D. Allison[5], Scott J. Bolton[6], Ravit Helled[7], Steven Levin[8], Cheng Li[1], Jonathan I. Lunine[9], Yamila Miguel[10], Glenn S. Orton[8], J. Hunter Waite[6], and Paul Withers[11]

[1] Department of Climate and Space Sciences and Engineering, University of Michigan, Ann Arbor, MI 48109-2143, USA; atreya@umich.edu
[2] School of Electrical and Computer Engineering, Georgia Institute of Technology, Atlanta, GA 30332-0250, USA
[3] School of Physics and Astronomy, University of Leicester, Leicester, LE1 7RH, UK
[4] Université Côte d'Azur, Observatoire de la Côte d'Azur, Laboratoire Lagrange, CNRS, Nice F-06304, France
[5] Williams College—Hopkins Observatory, Williamstown, MA 01267, USA
[6] Southwest Research Institute, San Antonio, TX 78238, USA
[7] Institute for Computational Science, Center for Theoretical Astrophysics & Cosmology, University of Zurich, CH-8057 Zurich, Switzerland
[8] Jet Propulsion Laboratory, California Institute of Technology, Pasadena, CA 91109, USA
[9] Astronomy Department, Cornell University, Ithaca, NY 14853, USA
[10] Leiden Observatory, University of Leiden, Niels Bohrweg 2, Leiden 2333CA, The Netherlands and SRON Netherlands Institute for Space Research, Niels Bohrweg 4, Leiden 2333CA, The Netherlands
[11] Astronomy Department, Boston University, Boston, MA 02215, USA
*Received 2022 March 20; revised 2022 April 19; accepted 2022 April 20; published 2022 July 14*

## Abstract

The thermal structure of planetary atmospheres is an essential input for predicting and retrieving the distribution of gases and aerosols, as well as the bulk chemical abundances. In the case of Jupiter, the temperature at a reference level—generally taken at 1 bar—serves as the anchor in models used to derive the planet's interior structure and composition. Most models assume the temperature measured by the Galileo probe. However, those data correspond to a single location, an unusually clear, dry region, affected by local atmospheric dynamics. On the other hand, the Voyager radio occultation observations cover a wider range of latitudes, longitudes, and times. The Voyager retrievals were based on atmospheric composition and radio refractivity data that require updating and were never properly tabulated; the few existing tabulations are incomplete and ambiguous. Here we present a systematic electronic digitization of all available temperature profiles from Voyager, followed by their reanalysis, employing currently accepted values of the abundances and radio refractivities of atmospheric species. We find the corrected temperature at the 1 bar level to be up to 4 K greater than the previously published values, i.e., $170.3 \pm 3.8$ K at 12°S (Voyager 1 ingress) and $167.3 \pm 3.8$ K at 0°N (Voyager 1 egress). This is to be compared with the Galileo probe value of $166.1 \pm 0.8$ K at the edge of an unusual feature at 6°.57N. Altogether, this suggests that Jupiter's tropospheric temperatures may vary spatially by up to 7 K between 7°N and 12°S.

*Unified Astronomy Thesaurus concepts:* Radio occultation (1351); Atmospheric refraction (115); Jupiter (873); Atmospheric composition (2120); Solar system formation (1530)

## 1. Introduction

The thermal structure of Jupiter's atmosphere is an important parameter controlled by numerous interlinked factors: the outward transport of internal energy (see Atreya 1986; Li et al. 2018); the latent heat released from the condensation of volatiles such as water, ammonium hydrosulfide, and ammonia that undergo phase transitions in the atmosphere (see Atreya et al. 1999); the radiative balance between short-wave solar heating and long-wave thermal emission (see Ingersoll et al. 2004; Moses et al. 2004); auroral heating (see Waite et al. 1983; Bougher et al. 2005; Sinclair et al. 2019); and dynamics (see Schneider & Liu 2009) in the stratosphere and the upper atmosphere. The thermal structure is used extensively in models of cloud formation, photochemistry, retrieval of elemental abundances and isotopic composition from observations (see Atreya 1986; Yung & Demore 1999; Atreya et al. 2019; Li et al. 2020), and vertical extrapolation of cloud-level winds (see Ingersoll et al. 2004; Fletcher et al. 2016).

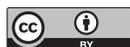

Moreover, the temperature at some reference level, generally 1 bar, serves as a critical anchor or tie point in models for retrieving the atmospheric composition (see Li et al. 2017, 2020) and thus affects the derived water abundance, which is a key constraint for Jupiter's formation models (see Owen et al. 1998, 1999; Owen & Encrenaz 2006). It is also a key parameter for interior models (see Wahl et al. 2017; Guillot et al. 2018), serving as the outer boundary condition for the temperature versus pressure in the deep interior (see Stevenson 2020).

Hence, many techniques—including infrared remote sensing; radio, solar, and stellar occultations; and in situ measurement—have been employed since the 1970s to provide robust determinations of the atmospheric temperature profile (see Kliore et al. 1976; Broadfoot et al. 1977; Elliot 1979; Festou et al. 1981; Conrath et al. 1998; Seiff et al. 1998; Irwin 2003). However, only the Voyager radio occultations and the Galileo probe direct measurements covered a pressure range that includes the 1 bar level or deeper. Infrared remote sensing, utilizing the collision-induced molecular hydrogen continuum in the 15–200 $\mu$m range, also has sensitivity to temperatures but with degeneracies associated with aerosol opacity limiting the accuracy of retrievals at those pressures (see Conrath et al. 1998). While the Galileo probe made highly precise measurements from nanobars to the 22 bar level (Seiff et al. 1998) on 1995 December 7, they correspond to the single location of the probe entry site (6.57°N, 5.02°W), which turned out to be at the edge of a large, unusually dry and clear region known as a 5 μm hot spot (Orton et al. 1998). Radio occultations are an additional resource for determining the temperature structure that cover a wider range of latitudes, longitudes, and times of observations, with a narrower extent from 1 mb to the 1 bar level.

**Table 1.**
Voyager Radio Occultation Observations

| Voyager (V) | Date | Planetographic Latitude | Pressure Range (mb) |
| --- | --- | --- | --- |
| V1 ingress | 1979 Mar 5 | 12°S | 1–1000 |
| V1 egress | 1979 Mar 5 | 0° | 1–1000 |
| V2 ingress | 1979 Jul 10 | 70°S | NA (see text) |
| V2 egress | 1979 Jul 10 | 60°S | 1–60 (see text) |

The first radio occultation observations at Jupiter were carried out by Pioneer 10 and 11 in 1973 December and 1974 December, respectively (Kliore et al. 1976). However, uncertainty in the center of refraction, the choice of initial temperature, and the spacecraft oscillator drifts resulted in relatively large error bars of up to tens of degrees in the derived temperatures over the pressure levels probed. Moreover, none of the occultations penetrated down to the important 1 bar pressure level, the highest pressure being the Pioneer 11 exit occultation reaching 250 mb thanks to being closer to the limb. In spite of their limitations, the Pioneer observations were remarkable because they demonstrated the power of the radio occultation technique in determining the properties of the neutral and ionized atmospheres of the giant planets.

The next radio occultation observations of Jupiter, made by the Voyager 1 and 2 spacecraft in 1979 (Table 1), provided mixed results but were more successful than Pioneer (Eshleman et al. 1979). The Voyager 1 flyby provided temperature data from 1 mb down to the 1 bar level, whereas data from the Voyager 2 flyby were much more limited in depth. No Voyager 2 ingress data were reported, and only partial information from 1 to 60 mb was obtained from the Voyager 2 egress. Potential causes for this lack of data could be due to problems faced while tracking Voyager 2, the shallowness of the occultation due to the spacecraft trajectory, greater spacecraft-to-limb distance, and multipath propagation (Eshleman et al. 1979; Hinson et al. 1998). Moreover, the $S$-band (2.3 GHz) signal was lost for much of the entry occultation, and the $X$-band (8.4 GHz) had many dropouts (Eshleman et al. 1979). Thus, the most valuable data set was from Voyager 1, but this work reassesses both the neutral temperatures derived from Voyager 1 and the Voyager 2 egress radio occultations.

Radio occultation experiments determine the atmospheric radial refractivity profile from which relative pressure ($p$) and temperature ($T$) profiles can be derived. The absolute values of $p$ and $T$ can be obtained if the mean molecular mass ($M$) or $T$ or $p$ are known independently at a reference level (Eshleman 1973). The basic principle of the temperature and pressure determination from radio occultation is outlined below.

The atmospheric refractivity is related to the total atmospheric number density and the refractive volume by the following expression:

$$\nu = \mu - 1 = n\kappa_{\mathrm{av}} = n\sum_j \kappa_j n_j/n, \quad (1)$$

where $\mu$ is the refractive index, $n$ is the total atmospheric number density, and $\kappa$ is the refractive volume. For an atmosphere composed of a mixture of constituents, $\kappa_{av}$ is the summation of the $\kappa_j$ of each component, $j$, multiplied by their volume mixing ratios, $n_j/n$. The refractive volume of gases relevant to the radio occultation observations can be found in Withers & Moore (2020).

Once the density is determined, the atmospheric pressure, $p$, can be calculated using the hydrostatic equation,

$$dp/dz = -g(z)\rho(z), \quad (2)$$

and

$$\rho = nM, \quad (3)$$

where $\rho$ is the mass density, $M$ is the molecular mass, and $g$ is gravity. The temperature can then be obtained from the pressure using ideal gas law,

$$p = nkT, \quad (4)$$

where $k$ is the Boltzmann constant.

Ideal gas is a reasonable assumption down to the 1 bar limit of the radio occultation data at Jupiter, but it fails (for example) for Titan given the low temperatures and high densities (Lindal et al. 1983).

Implied in the above derivation of $p$ and $T$ (Equations (1)–(4)) is the assumption that the chemical composition of the atmosphere and the proportions of its components, hence the mean molecular mass, are known independently. This is true only for simple atmospheres with spectroscopically active components or for Jupiter given the Galileo probe measurements.

At the time of the Voyager measurements, very little was known about the abundance of gases in Jupiter's atmosphere; hence, a gas mixture comprising only hydrogen and helium was considered, with their volume mixing ratios taken at 89% $H_2$ and 11% He, which was the best determination at that time, derived from a combination of Pioneer 10 and 11 Infrared Radiometer results (Orton & Ingersoll 1976) and early results from the Voyager IRIS experiment (Hanel et al. 1979). The Galileo probe provided far more precise measurements of the He and $H_2$ volume mixing ratios, as well as the abundances of gases that are likely to contribute to the mean molecular mass and radio refraction. In addition, newer laboratory measurements have considerably improved on the values of the refractive indices of these gases. Considering the importance of the Voyager radio occultation data to present and future applications, such as the radio occultation observations of Jupiter planned from the Juno spacecraft, it is timely to update the Voyager radio occultation data on temperatures in the neutral atmosphere. Therefore, we present here such an update, beginning with a systematic digitization of all available Voyager radio occultation profiles, and then apply the corrections to the temperatures and pressures using the current composition and refractivity values, followed by the first comprehensive tabulation of the new temperature profiles, and,

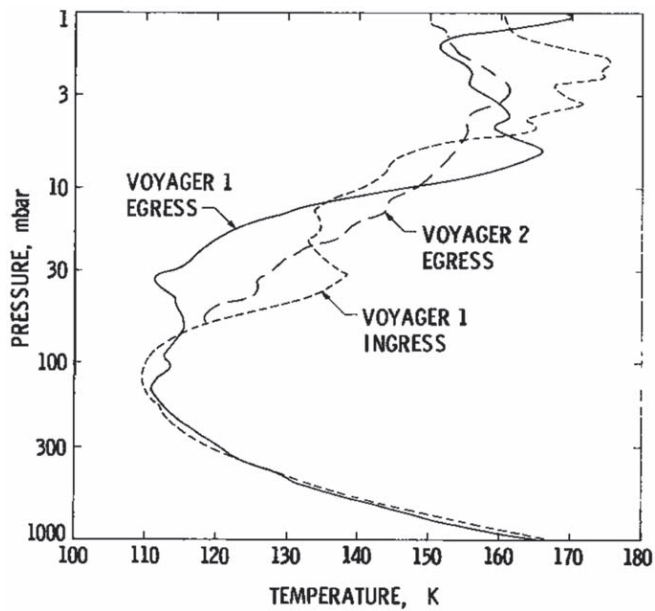

**Figure 1.** Temperature profile of Jupiter derived from radio occultation observations by Voyager 1 ingress and egress and Voyager 2 egress. Data for Voyager 2 ingress are unavailable (see text).

finally, a comparison with the infrared spectroscopic retrievals for context.

## 2. Methodology

### 2.1. Digitization

The raw data of the radio science observations from the Voyager missions (specifically for the Jupiter flybys) were never archived (confirmed by Dr. Edwin Bell of NSSDC), and the tabulation of temperatures from radio occultations was either missing, incomplete, or misleading in the few instances where it did exist. Therefore, it was necessary to first reconstruct the occultations based on published literature. This required digitization of all available Voyager radio occultation temperature profile plots contained in the figures of the original paper (Lindal et al. 1981). Corrections to both temperatures and pressures were then applied based on the latest available data on the abundances of gases relevant to the radio occultation altitude range and the laboratory measurements of the radio refractivities of those gases.

Using an electronic plot digitization tool,[12] temperature versus pressure were read off the Voyager 1 ingress, Voyager 1 egress, and Voyager 2 egress plots shown in Figure 3 of the original paper of Lindal et al. (1981; reproduced here as Figure 1).

A comparison of the digitized data against any previously available compilations was also carried out. The Planetary Data System (PDS) contains a partial tabulation of the hand-digitized values for Voyager 1 egress[13] with a reference to Lindal et al. (1981). However, it stops well short of the critical 1 bar level.

In addition to PDS, a listing for Jupiter's temperatures is found in Table 1 of Lindal (1992) along with a plot (Figure 10) on Voyager radio occultations at Neptune. However, the temperatures in the table and the corresponding plot for Jupiter in Figure 10 of that paper do not match. The figure's caption states 12°S, which would imply that the Jupiter plot is for Voyager 1 ingress, but that is misleading. A comparison of our own table of data (Section 4 of the current paper) digitized from the original plot for Voyager 1 ingress (Lindal et al. 1981) with our digitization of the Lindal (1992) Figure 10 Jupiter plot shows a discrepancy of ∼8 K in temperature at 1 mb, ∼2 K at 500 mb, and ∼3 K at ∼900 mb, which means that the Jupiter plot in Figure 10 of Lindal (1992) does not correspond to the Voyager 1 ingress profile. However, a visual inspection of that plot with the original plots of all Voyager profiles (Figure 3 of Lindal et al. 1981), as well as a comparison with all radio occultation temperature profiles we have digitized (Section 4 of the current paper) and the partial PDS tabulation, suggest that the Jupiter plot in Figure 10 of Lindal (1992) and the associated table most closely resemble the Voyager 1 egress profile. That is further confirmed by overlaying the plots we made of the PDS table entries, the Lindal (1992) table entries, and our own plot of the values digitized from the original paper (Lindal et al. 1981). We find a temperature of 165.2 K at 1 bar from digitizing the Voyager 1 egress plot of Lindal et al. (1981), which is within 0.2 K of the value listed in Lindal (1992). At other pressures, we also find a good agreement of between 0 and 0.3 K in the temperatures.

The close match between our $T$–$p$ values digitized from the original Voyager 1 egress plot of Lindal et al. (1981) and the tabulation in Lindal (1992), as well as the partial listing in PDS for the same, attests to the high fidelity (within 0–0.3 K) of the electronic digitization tool we have used. It also gives us confidence in the temperature profiles we digitized using that same tool on Voyager 1 ingress and Voyager 2 egress plots for which no other independent tabulations are available. With the digitization complete, we can now proceed to updating the profiles using the current information on the atmospheric composition and the refractivities of the relevant gases.

### 2.2. Temperature and Pressure Correction Factors

Corrections to the temperature versus pressure ($T$–$p$) data digitized from the plots in Lindal et al. (1981) were applied next. The temperature correction factor (TCF) is simply the ratio of the mean molecular mass of the atmosphere before and after updating the atmospheric composition, assuming ideal gases,

$$\text{TCF} = M\text{ (new)}/M\text{ (old)}, \quad (5)$$

where $M$(new) is the mean molecular mass after updating the atmospheric composition, and $M$(old) is the one used in the original work of Lindal et al. (1981).

The pressure correction factor (PCF) depends on both the change in the average refractivity of the atmosphere and the mean molecular mass,

$$\text{PCF} = [M\text{ (new)}/M\text{ (old)}] \times [\nu_{\text{av}}\text{ (old)}/\nu_{\text{av}}\text{ (new)}], \quad (6)$$

where $\nu_{\text{av}}$(new) is the weighted average refractivity of the new gas mixture after updating the composition, and $\nu_{\text{av}}$(old) is the one used in the original work of Lindal et al. (1981), which included only He and $H_2$. Equations (5) and (6) for TCF and PCF follow the approach outlined in Lindal et al. (1981) for scaling the temperatures and pressures for different compositions and refractivities assuming hydrostatic equilibrium and

---

[12] https://apps.automeris.io/wpd/
[13] https://pds-atmospheres.nmsu.edu/data_and_services/atmospheres_data/Voyager/VGR%201%20Egress%20hand%20digitized%20file.txt

Table 2
Volume Mixing Ratios of Gases in the Atmosphere of Jupiter[a]

| Cases | Hydrogen | Helium | Methane | Neon | Argon | Phosphine |
|---|---|---|---|---|---|---|
| Original[b] | 0.89 | 0.11 | 0 | 0 | 0 | 0 |
| Current[c] | 0.8623008 | 0.1356166 | 0.0020437 | 0.0000214 | 0.0000157 | 0.000001863 |

**Notes.**
[a] Von Zahn et al. (1998) for He/H, Wong et al. (2004) for methane, Mahaffy et al. (2000) for neon and argon, and Fletcher et al. (2009) for phosphine.
[b] Original refers to Lindal et al. (1981).
[c] Current refers to values used in this paper.

Table 3
Radio Refractivity ($\nu \times 10^6$) of Individual Gases in the Jovian Atmosphere[a]

| Cases | Hydrogen | Helium | Methane | Neon | Argon | Phosphine |
|---|---|---|---|---|---|---|
| Original[b] | 136 | 35 | 0 | 0 | 0 | 0 |
| Current[c] | 135.77 | 34.51 | 437.98 | 66.01 | 275.34 | 1267.52 |

**Notes.**
[a] Lindal et al. (1981) used the values from Essen (1953) for the only gases considered, He and $H_2$, which have since been updated with the improved laboratory measurements (Newell & Baird 1965) used in current paper. For the other gases included in this paper, the sources of the refractivity values at the radio frequency closest to the Voyager radio occultation frequencies are Asmar (2022) for methane, webtool for neon and argon (Polyanskiy 2008, plotting the data from Börzsönyi et al. 2008), and Mohammed & Steffes (2003) for phosphine and ammonia.
[b] Original refers to Lindal et al. (1981).
[c] Current refers to values used in this paper (references above).

negligible horizontal temperature gradients in the regions probed by the Voyager radio occultations.

### 3. Inputs and Resulting Correction Factors for Temperature and Pressure

#### 3.1. Atmospheric Composition

We used updated values derived from the Galileo probe measurements (Atreya et al. 2020). We substituted the very precise Galileo probe values for the He/$H_2$ ratio from the early results assumed by Lindal et al. (1981), with a larger proportion of helium than that employed previously. As noted above, we also included the abundances of other gases measured by the Galileo probe that are likely to contribute to the mean molecular mass and refractivity besides $H_2$ and He. Those gases include methane, argon, neon, phosphine, and ammonia ($CH_4$, Ar, Ne, $PH_3$, and $NH_3$, respectively). The abundances of all other gases are too low to make any significant impact on the derived temperature (Atreya et al. 2020). With the exception of $PH_3$ and $NH_3$, the constituents considered are expected to be well mixed in the 1–1000 mb pressure region of the radio occultation measurements, which is well below Jupiter's homopause, considering that they are noncondensable and nonreactive in this pressure range. Both $PH_3$ and $NH_3$ vary with altitude and latitude. However, even the maximum observed abundance of $PH_3$ is too low to significantly affect the mean molecular mass, and its effect on the temperature is less than 0.001 K at the 1 bar pressure level and much less at higher elevations, as discussed later. On the other hand, $NH_3$ can have a moderate effect on the mean molecular mass and, subsequently, the temperature. Considering that $NH_3$ varies with latitude (Li et al. 2017) and also undergoes condensation at 0.7 bar (Atreya et al. 1999), we treat it separately. The impact of ammonia on temperature is discussed in Section 4. Except for $NH_3$, it is thus reasonable to assume that for all practical purposes, the mean molecular mass remains constant throughout the radio occultation vertical regime for the neutral atmosphere. Water vapor and hydrogen sulfide ($H_2O$ and $H_2S$, respectively) are excluded because they are removed by condensation much deeper in the atmosphere, well below 1 bar (Atreya et al. 1999), the deepest level probed by the Voyager radio occultations at Jupiter. The current mixing ratios of all relevant constituents are listed in Table 2 along with the old values (only $H_2$ and He) used in the original publication (Lindal et al. 1981).

#### 3.2. Refractivity

Lindal et al. (1981) considered only He and $H_2$, using the values of refractivities available then. Improved values of the refractivities of He and $H_2$ are available, and other gases not previously considered make substantial contributions to the refraction of radio signal in the atmosphere. Table 3 lists the refractivity values for all gases included in this paper at the radio frequency closest to the Voyager radio occultation frequencies. In certain instances, the information is available as a refractive index ($\mu$), which can be converted to atmospheric refractivity ($\nu$) by using the relationship $\nu = \mu - 1$. A constant value of density-normalized refractivity was assumed for the different species, since laboratory measurements found that the refractivity of $H_2$, He, and $CH_4$ does not change with changes in pressure, temperature, or density (Spilker 1990). Other constituents (such as $PH_3$) have such a small contribution that they would not make any significant impact.

#### 3.3. Mean Molecular Mass and Weighted Average Refractivity

Based on the values in Tables 2 and 3, the mean molecular mass, average refractivity, and correction factors to be applied to temperature and pressure for updating the original radio occultation data were calculated. They are summarized in Table 4.

**Table 4**
Mean Molecular Mass ($M$) and Refractivity ($\nu_{av}$) of the Atmosphere along with the TCF and PCF

| Cases | $M$ (gm mole$^{-1}$) | $\nu_{av}$ | TCF | PCF |
|---|---|---|---|---|
| Original[a] | 2.2345 | 124.89 | NA | NA |
| Current[b] | 2.3151 | 122.66 | 1.03604 | 1.05489 |

**Notes.**
[a] Original refers to Lindal et al. (1981).
[b] Current refers to values used in this paper.

**Table 5**
Temperatures at 1 bar Pressure for Voyager 1

| Cases | V1 Ingress $T$ (K) | V1 Egress $T$ (K) |
|---|---|---|
| Original[a]: He+H$_2$ | 166.65 | 165.21 |
| Updated[b]: He+H$_2$ | 168.52 | 166.00 |
| Updated[b]: He+H$_2$+CH$_4$ | 170.21 | 167.29 |
| Updated[b]: He+H$_2$+CH$_4$+Ar+Ne+PH$_3$ | 170.28 | 167.32 |

**Notes.**
[a] Original refers to Lindal et al. (1981).
[b] Updated values are those calculated in this paper.

**Table 6**
Impact of NH$_3$ on the 1 bar Temperature for Voyager 1 Ingress

| NH$_3$ Conc. | PCF | TCF | $T$ at 1 bar Pressure |
|---|---|---|---|
| 0 ppmv | 1.05489 | 1.03604 | 170.28 |
| 100 ppmv | 1.05433 | 1.03666 | 170.39 |
| 200 ppmv | 1.05380 | 1.03723 | 170.53 |
| 300 ppmv | 1.05329 | 1.03779 | 170.64 |
| 400 ppmv | 1.05277 | 1.03836 | 170.76 |

## 4. Results

### 4.1. Temperature Profiles

After applying PCF and TCF from Table 4 to the original $T$–$p$ data of Lindal et al. (1981) digitized here, updated temperature versus pressure values were obtained for the three Voyager radio occultation observations. Table 5 summarizes the updated temperature results at the 1 bar pressure level for Voyager 1 ingress and egress (the Voyager 2 data stopped short of 1 bar) for different atmospheric compositions in a sequential order. The values listed in Table 5 are derived excluding the condensable ammonia, which is treated as discussed below, and its effect on temperature is summarized in Table 6. A complete listing of updated temperatures at all levels probed by Voyager 1 and 2 is provided in Tables 7 (V1 ingress), 8 (V1 egress), and 9 (V2 egress).

As expected, revision of the mean molecular mass makes the biggest change to temperature. For example, for the case with He and H$_2$ only, as in the original paper (Lindal et al. 1981), the new temperature at the 1 bar level, increases by approximately 1.9 K for Voyager 1 ingress and 0.9 K for Voyager 1 egress. The increased molecular mass accounts for most of the change, whereas the improved refractivity data contributes little. Addition of methane increases the temperature by another 1.7 K for Voyager 1 ingress and 1.4 K for Voyager 1 egress, again mostly because of increased molecular mass. Other gases considered in this study (Ar, Ne, and PH$_3$) do not impact the temperatures significantly because of their low abundances. For example, Ar and Ne together account for 0.07 K increase in the case of Voyager 1 ingress and 0.03 K for Voyager 1 egress, whereas even with the maximum observed abundance of PH$_3$ (Fletcher et al. 2009) the effect is ~0.005 K on the overall increase in temperature at 1 bar.

With the inclusion of all gases relevant to the Voyager radio occultation altitudes except ammonia (Table 5), we can now assess their overall impact on the 1 bar temperature. Compared with the originally published results (Lindal et al. 1981), the largest change is found for Voyager 1 ingress (12°S), with an increase of 3.6 K at the 1 bar level (and far greater at higher pressures), based on the digitized values of current paper, and 5.2 K relative to the value stated in the original publication (165 K). The updated values for Voyager 1 egress (0°) are greater by 2 K compared with the original value (Table 5).

Ammonia can also have a moderate effect on the temperature. The Juno observations show that over much of the planet, NH$_3$ is greatly depleted below its condensation level at 0.7 bar (Bolton et al. 2017; Ingersoll et al. 2017; Li et al. 2017). However, in a narrow latitude region near the equator, it is found to have a fairly constant mixing ratio with altitude, with an abundance of 351 ppmv, and then dropping to values as low as ~100 ppmv or even lower elsewhere (Li et al. 2017). At its maximum concentration, ammonia can result in an increase of 0.5 K in temperature between 0.7 and 1 bar pressure for the Voyager 1 ingress, and the increase of a similar magnitude for the Voyager 1 egress. The effect is negligible above the condensation level of ammonia (0.7 bar) because of its low saturation vapor pressure and loss due to photolysis. The specific contribution from different abundances of ammonia spanning the range found from Juno can be found in Table 6. The pressure and temperature correction factors (PCF and TCF) given in Table 6 include NH$_3$ together with all other gases (He+H$_2$+CH$_4$+Ar+Ne+PH$_3$).

### 4.2. Uncertainties

The uncertainty on the 1 bar temperature was estimated by Lindal et al. (1981) to be ±5 K, including the poorly known value of He/H$_2$, as well as uncertainties in the ephemeris and noise. In order to recalculate this uncertainty with the new information available, we proceed as follows. We first estimate the part of the original uncertainty that is not due to the uncertain helium to hydrogen ratio. To be able to compare the measurements at the two latitudes, we then add uncertainties resulting from new opacities that could differ as a function of latitude and time. We finally add all other sources of known uncertainties.

The uncertainty on the He/H$_2$ ratio considered by Lindal et al. (1981), 0.11 ± 0.03, amounts to a relative uncertainty on the mean molecular mass of 2.2%, implying a $T(z)$ relative uncertainty of 2.2%. However, because we are interested in the $T(p)$ value, we must account for the fact that $p(z)$ also changes. Using Table 7 hereafter, we see that from the old to the new analysis, $T(z)$ changes by 3.6% (e.g., 166.65 to 172.66 K using the last points in Table 7), whereas the corresponding change in $T_{1bar}$ is 2.2% (i.e., 166.65 to 170.28 K). Therefore, we estimate that the relative uncertainty on $T_{1bar}$ linked to the He/H$_2$ ratio in 1981 was 0.022 × 0.022/0.036 = 1.3%, implying that the core uncertainty was 3%–1.3% = 1.7%, amounting to a ±2.9 K uncertainty in the 1 bar temperature. Because the ammonia abundance varies significantly as a function of latitude (Li et al. 2017; Bolton et al. 2017) and leads to a change in $T_{1bar}$ between 0 and 0.5 K (see Table 6), we need to add this source of uncertainty, leading to a value of ±3.2 K. (The uncertainty on $PH_3$ abundance—also variable as a function of latitude—may be neglected, as discussed in Section 4.1.)

Table 7
Voyager 1 Ingress T–p: Original (Lindal et al. 1981) and Updated (This Paper)

| Original (Only $H_2$ at 89% and He at 11%) | | Updated ($H_2$ + He + $CH_4$ + Ne + Ar + $PH_3$) | | Original (Only $H_2$ at 89% and He at 11%) | | Updated ($H_2$ + He + $CH_4$ + Ne + Ar + $PH_3$) | |
|---|---|---|---|---|---|---|---|
| P (mb) | T (K) | P (mb) | T (K) | P (mb) | T (K) | P (mb) | T (K) |
| 1.07 | 160.57 | 1.13 | 166.36 | 18.41 | 134.03 | 19.42 | 138.86 |
| 1.17 | 160.98 | 1.23 | 166.78 | 19.32 | 133.52 | 20.38 | 138.33 |
| 1.22 | 161.19 | 1.29 | 167.00 | 20.26 | 133.11 | 21.37 | 137.91 |
| 1.32 | 162.01 | 1.39 | 167.85 | 21.05 | 133.32 | 22.21 | 138.12 |
| 1.40 | 163.35 | 1.48 | 169.24 | 22.28 | 133.93 | 23.50 | 138.76 |
| 1.44 | 164.17 | 1.52 | 170.09 | 24.74 | 134.76 | 26.10 | 139.62 |
| 1.46 | 165.10 | 1.54 | 171.05 | 26.68 | 135.89 | 28.14 | 140.79 |
| 1.64 | 170.33 | 1.73 | 176.47 | 27.97 | 136.71 | 29.51 | 141.64 |
| 1.65 | 171.05 | 1.74 | 177.21 | 32.57 | 137.54 | 34.36 | 142.50 |
| 1.68 | 171.87 | 1.77 | 178.06 | 35.16 | 136.93 | 37.09 | 141.86 |
| 1.75 | 173.72 | 1.85 | 179.98 | 37.61 | 135.91 | 39.67 | 140.81 |
| 1.80 | 174.44 | 1.90 | 180.73 | 38.72 | 134.88 | 40.85 | 139.74 |
| 1.83 | 175.05 | 1.93 | 181.36 | 41.84 | 132.22 | 44.14 | 136.99 |
| 1.87 | 175.26 | 1.97 | 181.58 | 45.21 | 129.76 | 47.69 | 134.44 |
| 1.94 | 175.26 | 2.05 | 181.58 | 47.00 | 128.22 | 49.58 | 132.84 |
| 2.20 | 174.34 | 2.32 | 180.62 | 49.34 | 126.28 | 52.05 | 130.83 |
| 2.28 | 174.45 | 2.41 | 180.74 | 51.79 | 124.74 | 54.63 | 129.24 |
| 2.37 | 173.94 | 2.50 | 180.21 | 54.41 | 121.36 | 57.40 | 125.73 |
| 2.42 | 173.12 | 2.55 | 179.36 | 59.93 | 118.49 | 63.22 | 122.76 |
| 2.51 | 170.56 | 2.65 | 176.71 | 62.90 | 116.95 | 66.35 | 121.16 |
| 2.54 | 169.63 | 2.68 | 175.74 | 73.39 | 113.47 | 77.42 | 117.56 |
| 2.59 | 167.68 | 2.73 | 173.72 | 80.02 | 112.15 | 84.41 | 116.19 |
| 2.77 | 168.30 | 2.92 | 174.37 | 93.28 | 110.41 | 98.40 | 114.39 |
| 2.90 | 168.82 | 3.06 | 174.90 | 124.22 | 109.70 | 131.04 | 113.65 |
| 3.07 | 170.15 | 3.24 | 176.28 | 162.12 | 111.15 | 171.02 | 115.16 |
| 3.19 | 170.87 | 3.37 | 177.03 | 196.04 | 112.60 | 206.80 | 116.66 |
| 3.35 | 171.39 | 3.53 | 177.57 | 236.97 | 114.86 | 249.98 | 119.00 |
| 3.65 | 167.49 | 3.85 | 173.53 | 289.01 | 118.36 | 304.67 | 122.63 |
| 3.72 | 166.26 | 3.92 | 172.25 | 332.93 | 121.64 | 351.20 | 126.02 |
| 3.87 | 165.04 | 4.08 | 170.99 | 369.18 | 124.83 | 389.44 | 129.33 |
| 3.99 | 164.52 | 4.21 | 170.45 | 445.47 | 130.79 | 469.92 | 135.50 |
| 4.26 | 164.12 | 4.49 | 170.03 | 449.51 | 131.04 | 474.18 | 135.76 |
| 4.30 | 164.32 | 4.54 | 170.24 | 471.47 | 132.49 | 497.35 | 137.27 |
| 4.34 | 164.63 | 4.58 | 170.56 | 485.15 | 133.43 | 511.78 | 138.24 |
| 4.73 | 164.74 | 4.99 | 170.68 | 494.47 | 134.16 | 521.61 | 138.99 |
| 4.78 | 163.10 | 5.04 | 168.98 | 503.96 | 134.99 | 531.62 | 139.85 |
| 4.87 | 162.17 | 5.14 | 168.01 | 513.64 | 135.71 | 541.84 | 140.61 |
| 4.92 | 161.35 | 5.19 | 167.17 | 528.53 | 136.75 | 557.54 | 141.68 |
| 5.02 | 160.64 | 5.30 | 166.43 | 538.72 | 137.27 | 568.29 | 142.22 |
| 5.12 | 158.69 | 5.40 | 164.41 | 554.33 | 138.31 | 584.76 | 143.30 |
| 5.23 | 155.92 | 5.52 | 161.54 | 570.40 | 139.35 | 601.71 | 144.37 |
| 5.43 | 154.69 | 5.73 | 160.27 | 592.47 | 141.12 | 624.99 | 146.20 |
| 5.49 | 151.72 | 5.79 | 157.19 | 609.60 | 142.37 | 643.06 | 147.50 |
| 5.82 | 149.87 | 6.14 | 155.27 | 627.16 | 143.92 | 661.59 | 149.11 |
| 5.99 | 148.85 | 6.32 | 154.21 | 645.34 | 144.96 | 680.76 | 150.19 |
| 6.23 | 147.31 | 6.57 | 152.62 | 676.78 | 146.83 | 713.93 | 152.13 |
| 6.54 | 146.19 | 6.90 | 151.46 | 702.93 | 148.81 | 741.51 | 154.17 |
| 6.67 | 145.47 | 7.04 | 150.71 | 723.25 | 150.05 | 762.95 | 155.46 |
| 7.20 | 144.55 | 7.60 | 149.76 | 743.99 | 152.03 | 784.82 | 157.51 |
| 8.31 | 143.63 | 8.77 | 148.81 | 758.28 | 152.76 | 799.91 | 158.26 |
| 8.63 | 143.02 | 9.10 | 148.17 | 772.80 | 153.69 | 815.22 | 159.23 |
| 9.06 | 142.30 | 9.56 | 147.43 | 787.65 | 154.42 | 830.89 | 159.98 |
| 9.78 | 140.97 | 10.32 | 146.05 | 802.74 | 155.35 | 846.80 | 160.95 |
| 10.07 | 139.85 | 10.62 | 144.89 | 833.75 | 157.33 | 879.51 | 163.00 |
| 10.47 | 138.93 | 11.04 | 143.94 | 849.69 | 158.37 | 896.32 | 164.07 |
| 10.88 | 137.80 | 11.48 | 142.77 | 857.65 | 159.30 | 904.73 | 165.04 |
| 11.53 | 136.47 | 12.16 | 141.39 | 890.81 | 161.17 | 939.71 | 166.98 |
| 12.57 | 134.83 | 13.26 | 139.69 | 916.42 | 162.94 | 966.72 | 168.81 |
| 13.31 | 134.12 | 14.04 | 138.95 | 948.03 | 164.36 | 1000.07 | 170.28 |
| 14.10 | 134.12 | 14.87 | 138.95 | 966.12 | 165.30 | 1019.15 | 171.26 |
| 14.93 | 134.53 | 15.75 | 139.38 | 994.03 | 166.34 | 1048.59 | 172.33 |
| 15.80 | 134.84 | 16.67 | 139.70 | 1000.03 | 166.65 | 1054.92 | 172.66 |
| 17.55 | 134.44 | 18.51 | 139.29 | | | | |

Table 8
Voyager 1 Egress $T$–$p$: Original (Lindal et al. 1981) and Updated (This Paper)

| Original (Only $H_2$ at 89% and He at 11%) | | Updated ($H_2$ + He + $CH_4$ + Ne + Ar + $PH_3$) | | Original (Only $H_2$ at 89% and He at 11%) | | Updated ($H_2$ + He + $CH_4$ + Ne + Ar + $PH_3$) | |
|---|---|---|---|---|---|---|---|
| $P$ (mb) | $T$ (K) | $P$ (mb) | $T$ (K) | $P$ (mb) | $T$ (K) | $P$ (mb) | $T$ (K) |
| 1.12 | 168.88 | 1.18 | 174.97 | 16.34 | 124.18 | 17.24 | 128.66 |
| 1.13 | 167.34 | 1.19 | 173.37 | 18.00 | 122.03 | 18.99 | 126.43 |
| 1.16 | 165.80 | 1.22 | 171.78 | 19.63 | 120.08 | 20.71 | 124.41 |
| 1.19 | 164.68 | 1.26 | 170.62 | 20.79 | 118.96 | 21.93 | 123.25 |
| 1.20 | 163.75 | 1.27 | 169.65 | 23.33 | 117.32 | 24.61 | 121.55 |
| 1.23 | 162.01 | 1.30 | 167.85 | 24.71 | 116.91 | 26.07 | 121.12 |
| 1.25 | 160.78 | 1.32 | 166.57 | 26.17 | 116.40 | 27.61 | 120.60 |
| 1.30 | 158.12 | 1.37 | 163.82 | 27.46 | 115.58 | 28.97 | 119.75 |
| 1.31 | 157.30 | 1.38 | 162.97 | 28.81 | 114.66 | 30.39 | 118.79 |
| 1.32 | 156.17 | 1.39 | 161.80 | 29.66 | 113.74 | 31.29 | 117.84 |
| 1.34 | 154.94 | 1.41 | 160.52 | 30.25 | 112.92 | 31.91 | 116.99 |
| 1.35 | 154.12 | 1.42 | 159.67 | 30.84 | 112.10 | 32.53 | 116.14 |
| 1.38 | 153.30 | 1.46 | 158.82 | 33.30 | 111.49 | 35.13 | 115.51 |
| 1.42 | 152.68 | 1.50 | 158.18 | 35.93 | 111.90 | 37.90 | 115.93 |
| 1.46 | 152.17 | 1.54 | 157.65 | 37.67 | 112.32 | 39.74 | 116.37 |
| 1.53 | 151.66 | 1.61 | 157.13 | 39.50 | 113.14 | 41.67 | 117.22 |
| 1.61 | 151.66 | 1.70 | 157.13 | 41.02 | 113.96 | 43.27 | 118.07 |
| 1.68 | 152.08 | 1.77 | 157.56 | 44.26 | 114.27 | 46.69 | 118.39 |
| 1.75 | 152.59 | 1.85 | 158.09 | 47.31 | 114.38 | 49.91 | 118.50 |
| 1.80 | 153.10 | 1.90 | 158.62 | 52.03 | 115.00 | 54.89 | 119.14 |
| 1.98 | 154.85 | 2.09 | 160.43 | 54.56 | 115.31 | 57.55 | 119.47 |
| 2.07 | 155.47 | 2.18 | 161.07 | 60.59 | 115.52 | 63.92 | 119.68 |
| 2.22 | 155.78 | 2.34 | 161.39 | 66.03 | 115.21 | 69.65 | 119.36 |
| 2.35 | 155.78 | 2.48 | 161.39 | 73.36 | 114.50 | 77.39 | 118.63 |
| 2.51 | 155.58 | 2.65 | 161.19 | 80.73 | 113.68 | 85.16 | 117.78 |
| 2.66 | 155.79 | 2.81 | 161.40 | 89.69 | 112.87 | 94.61 | 116.94 |
| 2.73 | 156.30 | 2.88 | 161.93 | 94.06 | 113.08 | 99.22 | 117.16 |
| 2.81 | 156.82 | 2.96 | 162.47 | 99.58 | 113.49 | 105.05 | 117.58 |
| 2.89 | 157.43 | 3.05 | 163.10 | 106.45 | 113.49 | 112.29 | 117.58 |
| 3.00 | 158.05 | 3.16 | 163.75 | 114.94 | 112.57 | 121.25 | 116.63 |
| 3.24 | 159.18 | 3.42 | 164.92 | 125.30 | 111.45 | 132.18 | 115.47 |
| 3.40 | 160.00 | 3.59 | 165.77 | 148.78 | 111.25 | 156.95 | 115.26 |
| 3.53 | 160.42 | 3.72 | 166.20 | 166.74 | 112.18 | 175.89 | 116.22 |
| 3.84 | 161.04 | 4.05 | 166.84 | 211.33 | 115.06 | 222.93 | 119.21 |
| 4.03 | 160.73 | 4.25 | 166.52 | 334.40 | 122.20 | 352.75 | 126.60 |
| 4.23 | 160.01 | 4.46 | 165.78 | 598.41 | 140.08 | 631.25 | 145.13 |
| 4.39 | 159.30 | 4.63 | 165.04 | 664.46 | 144.03 | 700.94 | 149.22 |
| 4.61 | 159.51 | 4.86 | 165.26 | 683.65 | 145.38 | 721.18 | 150.62 |
| 4.83 | 160.33 | 5.10 | 166.11 | 723.78 | 147.77 | 763.51 | 153.10 |
| 4.97 | 161.05 | 5.24 | 166.85 | 737.66 | 148.60 | 778.15 | 153.96 |
| 5.16 | 162.18 | 5.44 | 168.02 | 751.81 | 149.43 | 793.08 | 154.82 |
| 5.46 | 163.41 | 5.76 | 169.30 | 781.07 | 150.58 | 823.94 | 156.00 |
| 5.62 | 164.33 | 5.93 | 170.25 | 796.00 | 151.62 | 839.69 | 157.08 |
| 5.89 | 165.46 | 6.21 | 171.42 | 818.79 | 153.69 | 863.74 | 159.23 |
| 6.80 | 164.14 | 7.17 | 170.06 | 834.47 | 154.63 | 880.28 | 160.20 |
| 7.42 | 161.48 | 7.83 | 167.30 | 850.42 | 155.67 | 897.10 | 161.28 |
| 7.94 | 158.91 | 8.38 | 164.64 | 866.68 | 156.71 | 914.25 | 162.35 |
| 8.42 | 156.35 | 8.88 | 161.98 | 883.33 | 157.43 | 931.82 | 163.11 |
| 9.01 | 153.69 | 9.50 | 159.23 | 900.28 | 158.27 | 949.69 | 163.97 |
| 9.46 | 150.31 | 9.98 | 155.73 | 926.12 | 160.14 | 976.96 | 165.91 |
| 10.04 | 146.31 | 10.59 | 151.58 | 943.83 | 161.17 | 995.63 | 166.98 |
| 10.44 | 143.64 | 11.01 | 148.82 | 952.64 | 162.21 | 1004.94 | 168.06 |
| 11.18 | 139.85 | 11.79 | 144.89 | 970.92 | 163.04 | 1024.21 | 168.92 |
| 11.98 | 136.16 | 12.64 | 141.07 | 980.09 | 163.77 | 1033.89 | 169.67 |
| 12.70 | 132.88 | 13.40 | 137.67 | 989.31 | 164.60 | 1043.62 | 170.53 |
| 13.99 | 129.71 | 14.76 | 134.38 | 1000.06 | 165.21 | 1054.95 | 171.16 |
| 15.27 | 126.74 | 16.11 | 131.31 | | | | |

The systematic uncertainties due to the inclusion of new helium and methane abundances may be estimated as follows. The constraint on the helium abundance from the Galileo probe (von Zahn et al. 1998) is 10 times more accurate than the 1981 one; thus, its relative uncertainty is only 0.13%, or a ±0.2 K change. The relative uncertainty on methane abundance is 24% (Wong et al. 2004); thus, given its ∼1.7 K effect (Table 5), we estimate the corresponding uncertainty to ±0.4 K.

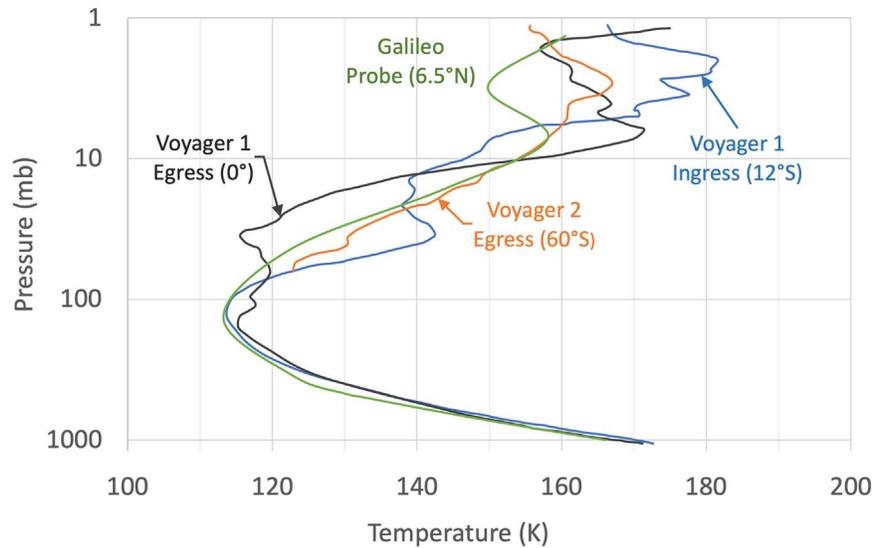

**Figure 2.** Updated temperature profiles of Jupiter derived from radio occultation observations by the Voyager 1 ingress and egress and the Voyager 2 egress, together with the Galileo probe measurements. Data for the Voyager 2 ingress are unavailable (see text).

Table 9
Voyager 2 Egress T–p: Original (Lindal et al. 1981) and Updated (This Paper)

| Original (Only $H_2$ at 89% and He at 11%) | | Updated ($H_2$ + He + $CH_4$ + Ne + Ar + $PH_3$) | | Original (Only $H_2$ at 89% and He at 11%) | | Updated ($H_2$ + He + $CH_4$ + Ne + Ar + $PH_3$) | |
|---|---|---|---|---|---|---|---|
| P (mb) | T (K) | P (mb) | T (K) | P (mb) | T (K) | P (mb) | T (K) |
| 1.08 | 150.21 | 1.14 | 155.62 | 7.03 | 152.34 | 7.42 | 157.83 |
| 1.15 | 150.21 | 1.21 | 155.62 | 7.67 | 151.43 | 8.09 | 156.89 |
| 1.20 | 150.83 | 1.27 | 156.27 | 8.44 | 150.30 | 8.90 | 155.72 |
| 1.22 | 151.55 | 1.29 | 157.01 | 8.69 | 149.89 | 9.17 | 155.29 |
| 1.29 | 152.17 | 1.36 | 157.65 | 9.11 | 149.18 | 9.61 | 154.56 |
| 1.36 | 152.58 | 1.43 | 158.08 | 10.03 | 147.85 | 10.58 | 153.18 |
| 1.46 | 152.89 | 1.54 | 158.40 | 10.63 | 146.52 | 11.21 | 151.80 |
| 1.54 | 153.20 | 1.62 | 158.72 | 10.84 | 145.90 | 11.44 | 151.16 |
| 1.59 | 153.51 | 1.68 | 159.04 | 11.16 | 145.29 | 11.77 | 150.53 |
| 1.73 | 154.23 | 1.82 | 159.79 | 11.82 | 144.47 | 12.47 | 149.68 |
| 1.78 | 154.54 | 1.88 | 160.11 | 12.52 | 143.86 | 13.21 | 149.04 |
| 1.83 | 155.26 | 1.93 | 160.86 | 14.04 | 143.04 | 14.81 | 148.20 |
| 1.87 | 155.57 | 1.97 | 161.18 | 15.02 | 140.79 | 15.84 | 145.86 |
| 1.94 | 156.49 | 2.05 | 162.13 | 15.92 | 139.66 | 16.79 | 144.69 |
| 2.01 | 157.21 | 2.12 | 162.88 | 17.02 | 138.74 | 17.95 | 143.74 |
| 2.11 | 157.93 | 2.23 | 163.62 | 18.20 | 137.93 | 19.20 | 142.90 |
| 2.21 | 158.75 | 2.33 | 164.47 | 19.85 | 136.60 | 20.94 | 141.52 |
| 2.28 | 159.37 | 2.41 | 165.11 | 20.24 | 135.78 | 21.35 | 140.67 |
| 2.36 | 159.78 | 2.49 | 165.54 | 20.64 | 134.75 | 21.77 | 139.61 |
| 2.43 | 160.30 | 2.56 | 166.08 | 23.17 | 131.99 | 24.44 | 136.75 |
| 2.50 | 160.61 | 2.64 | 166.40 | 24.31 | 131.07 | 25.64 | 135.79 |
| 2.55 | 160.81 | 2.69 | 166.61 | 25.02 | 130.55 | 26.39 | 135.26 |
| 2.67 | 161.12 | 2.82 | 166.93 | 26.01 | 129.74 | 27.44 | 134.42 |
| 2.86 | 161.12 | 3.02 | 166.93 | 27.29 | 128.92 | 28.79 | 133.57 |
| 3.24 | 159.90 | 3.42 | 165.66 | 30.04 | 127.28 | 31.69 | 131.87 |
| 3.36 | 159.29 | 3.54 | 165.03 | 31.22 | 126.77 | 32.93 | 131.34 |
| 3.50 | 158.47 | 3.69 | 164.18 | 32.75 | 126.26 | 34.55 | 130.81 |
| 3.60 | 156.83 | 3.80 | 162.48 | 34.68 | 125.85 | 36.58 | 130.39 |
| 3.71 | 156.01 | 3.91 | 161.63 | 38.52 | 125.86 | 40.63 | 130.40 |
| 3.93 | 155.40 | 4.15 | 161.00 | 41.19 | 124.94 | 43.45 | 129.44 |
| 4.20 | 155.30 | 4.43 | 160.90 | 42.40 | 124.22 | 44.73 | 128.70 |
| 4.57 | 155.20 | 4.82 | 160.79 | 43.24 | 123.19 | 45.61 | 127.63 |
| 4.98 | 155.10 | 5.25 | 160.69 | 45.40 | 121.25 | 47.89 | 125.62 |
| 5.43 | 154.49 | 5.73 | 160.06 | 48.56 | 119.92 | 51.23 | 124.24 |
| 5.70 | 154.18 | 6.01 | 159.74 | 51.94 | 118.90 | 54.79 | 123.19 |
| 6.27 | 153.26 | 6.61 | 158.78 | 58.24 | 118.59 | 61.44 | 122.86 |
| 6.64 | 152.75 | 7.00 | 158.26 | 59.81 | 118.60 | 63.09 | 122.87 |

Properly combining the estimated uncertainties would require a full reanalysis of the Voyager radio science data that is beyond the scope of the present paper. Therefore, in order to account for the correlated effects, we simply add the uncertainties linearly and estimate an uncertainty on the 1 bar temperature at each latitude of ±3.2 K and a global uncertainty of ±3.8 K. Following Lindal et al. (1981), we caution that possible systematic errors are not included in this analysis.

Now let us consider the uncertainty in the Galileo probe analysis, since it is not directly provided. Following Seiff et al. (1998; see also Magalhaes et al. 2002), the uncertainty in the temperature measurement increases from 0.1 K at 100 K to 1 K at 500 K, which amounts to a 0.25 K uncertainty for the measurement. The uncertainty on the pressure sensor at that level is estimated to be $\delta_p = 0.018$ bar (Table 4 of Seiff et al. 1998), yielding a corresponding temperature uncertainty of $\delta_T = \delta_p(dT/dp) \sim 0.75$ K. Adding the uncertainties quadratically yields a temperature value from the Galileo probe of $166.1 \pm 0.8$ K. Altogether, we derive temperatures of $170.3 \pm 3.2$ (±3.8 global) K at 12°S (Voyager 1 ingress) and $167.3 \pm 3.2$ (±3.8 global) K at 0°N (Voyager 1 egress). This is to be compared with the Galileo probe value at the edge of a 5 μm hot spot at 6°.57N, $166.1 \pm 0.8$ K.

Two approaches may be considered. (1) Traditionally, these values are considered as representative of one uniform temperature across the entire Jupiter disk. In this case, the Galileo probe value, which has by far the smallest uncertainty, should be chosen. However, we note that with our new analysis, there is now some inconsistency between the Galileo probe value and the 12°S Voyager 1 ingress values whose error bars do not overlap. Hence, this motivates a second approach. (2) These values may differ because of horizontal fluctuations (mostly as a function of latitude) of the temperature–pressure field. In this case, temperature fluctuations should be limited to between 1 and 7 K at latitudes between 7°N and 12°S.

## 5. Comparison to Spectroscopic Retrievals

Although Jupiter's vertical temperature structure can also be derived from infrared remote sensing, such studies are more sensitive to the upper troposphere and stratosphere than they are to the 1 bar level. Far-infrared remote sensing of the collision-induced hydrogen–helium continuum in the 15–200 μm region (see Conrath et al. 1998; Fletcher et al. 2017) does provide sensitivity to the spatial distributions of the 1 bar opacity. However, this opacity is shaped by the atmospheric temperature, the ratio of the spin isomers of hydrogen (e.g., the para-$H_2$ versus ortho-$H_2$ ratios), and the uncertain contribution of aerosol absorption from large tropospheric particles. We now investigate whether the updated $T(p)$ from the radio science experiment (RSS) can be made consistent with spectroscopic retrievals.

We first investigate the sensitivity of mid- and far-IR temperature retrievals to the updated $T(p)$ profiles using a profile with temperatures shifted to 175 K (encompassing the new temperature range of the present paper) as an initial assumption (prior) to rerun infrared spectral inversions with the NEMESIS suite of radiative transfer and spectral retrieval tools (Irwin et al. 2008). Retrievals of temperature were performed with (i) Cassini/CIRS 2.5 cm$^{-1}$ spectral resolution maps of Jupiter from 7 to 16 μm (ATMOS02A) as presented by Fletcher et al. (2016) and (ii) Voyager 1/IRIS inbound 4.3 cm$^{-1}$ spectral maps from 7 to 29 μm as presented by Fletcher et al. (2017). The former provides some limited sensitivity to temperatures, although this is degenerate with assumptions about the $NH_3$ and $PH_3$ vertical profiles in the 8–11 μm region. The latter provides more sensitivity to the deep temperature, albeit at a degraded spatial and spectral resolution. Both inversions used the latest estimates of collision-induced $H_2$ opacity and associated dimeric absorption (Fletcher et al. 2018).

In both sets of retrievals, the quality of the spectral fits remained unchanged between the 166 K original case and the 175 K extreme case, indicating that the new priors are perfectly acceptable for infrared observations. This is not surprising; thermal IR studies are typically suggested to be accurate above the $NH_3$ ice cloud at $p < 0.7$ bar, and the information content at 1 bar is low. The retrieved aerosol opacities increased somewhat to balance the warmer temperatures; if some way was found to break the temperature–aerosol degeneracy (e.g., by consideration of cloud reflectivity at shorter wavelengths), it might place stronger constraints on the deep temperature. The temperatures retrieved from Voyager/IRIS suggest that anything in the 160–175 (±2) K range is acceptable, with generally lower temperatures (160–167 K) at low latitudes and higher temperatures (167–175 K) beyond ±45° in each hemisphere. We caution that these estimates remain highly degenerate with assumptions about aerosol opacity. These estimates are consistent with a recent study of ground-based high-resolution spectroscopy from the TEXES instrument (Fletcher et al. 2020), which suggested that temperatures associated with 5 μm hot spots on Jupiter's north equatorial belt could be 2–3 K cooler than the Galileo-derived value of 166 K (Seiff et al. 1998), and that latitudinal variations of temperatures of 3–5 K could be consistent with the data within ±25° of the equator, albeit with ±2 K uncertainties at 1 bar (1σ).

As a second test, both the old and new Voyager 1/RSS ingress and egress $T(p)$ profiles were used as priors for Voyager/IRIS inversions, heavily constraining the temperatures to match the priors but allowing free variation in the upper troposphere and stratosphere. IRIS spectra between 280 and 700 cm$^{-1}$ were zonally averaged within ±2°.5 latitude of the ingress (12°S) and egress (0°) latitudes. This implicitly assumes zonal homogeneity, ignoring the effects of thermal waves and/or large vortices in the latitude domain. The IRIS spectra were fit to within ±0.2 K of the brightness temperature, and the aerosol and para-$H_2$ profiles were consistent with previous studies, irrespective of whether the old or new profiles were used, confirming the low sensitivity to the temperatures. The RSS profiles (both ingress and egress and old and new estimates) and retrieved IRIS profiles (using either the old or new $T(p)$ as priors) are shown in Figure 3(a), revealing that IRIS inversions can be made consistent with either the old or revised 1 bar temperatures without detriment to the overall retrieval quality. Figure 3(a) shows IRIS $T(p)$ estimates converging in the 80–600 mb region, where the spectra offer maximal constraint on the temperatures, but they diverge at higher pressures, where they are more strongly influenced by the choice of prior.

However, Figure 3(b) reveals differences between the IRIS and RSS profiles for both the new and old $T(p)$ priors. For the new $T(p)$priors, the 400–800 mb temperatures needed to be persistently cooler than the RSS profiles in order to reproduce the spectra. For the old $T(p)$ priors, the retrieved 400–800 mb temperatures were consistent with the old RSS profiles, but the 100–400 mb IRIS temperatures were warmer than the RSS profiles. In the range where IRIS spectra can constrain temperatures, it therefore seems difficult to impose consistency between the IRIS and RSS profiles, as first pointed out by Conrath & Gautier (2000). However, the inconsistency might be resolved by allowing parameters like para-$H_2$ and aerosols to vary or could be explained by the IRIS spectra and RSS profiles sampling different latitudes/longitudes with different spatial resolutions.

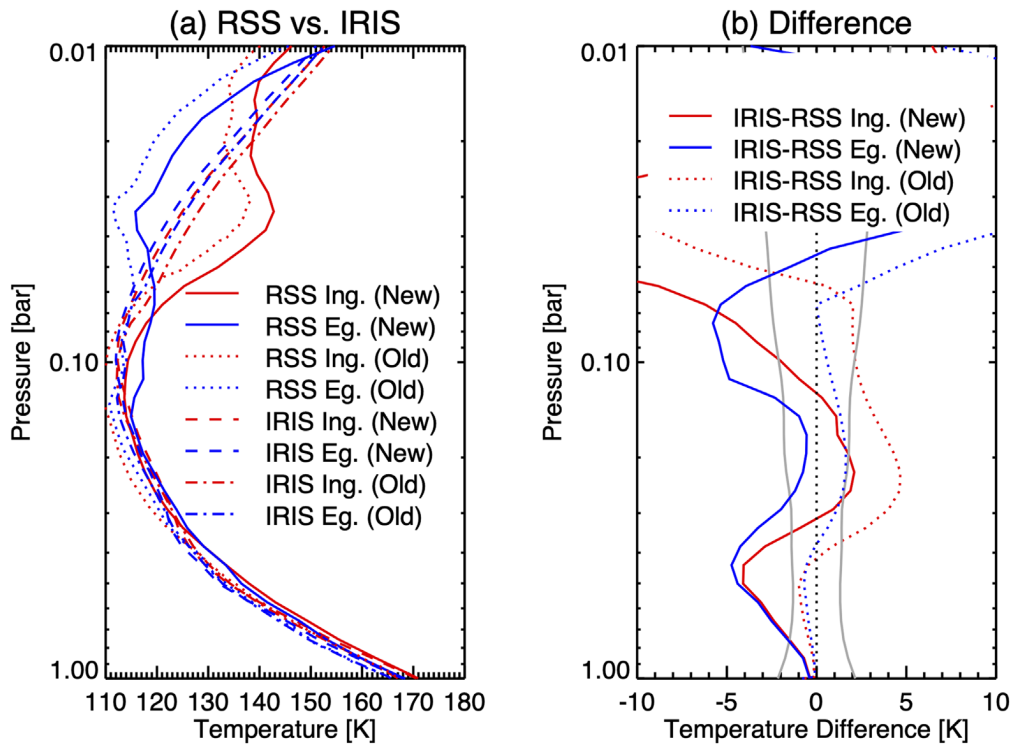

**Figure 3.** Comparison of Voyager 1 radio occultations (RSS) and IRIS retrieved temperatures. Ingress profiles are shown in red; egress profiles are shown in blue. The newly derived RSS profiles are solid lines, and the old versions are dotted lines. The profiles retrieved from IRIS zonally averaged spectra are shown as dashed (when the new $T(p)$ was used as a prior) and dotted–dashed (when the old $T(p)$ was used as a prior) lines. Uncertainties on the IRIS retrievals are shown as the gray lines in panel (b). Note how the IRIS retrievals converge in the 80–600 mb range, where they offer maximal constraint on the $T(p)$, but diverge at higher pressures, where they are more strongly influenced by the choice of prior.

These comparisons of the thermal infrared and RSS profiles are at the edge of what is currently possible, and extensive latitudinal coverage by the radio occultation observations planned for the Juno extended mission may enable more robust assessments of spatial variability in temperatures.

## 6. Discussion and Conclusions

The updated temperatures suggested by this reanalysis have repercussions for studies that critically depend on Jupiter's vertical temperature structure. It has often been implicitly assumed that the accurate Galileo probe measurement at 1 bar, $166.1 \pm 0.8$ K, is appropriate for the entire planet. However, modeling of the 5 $\mu$m hot-spot regions (Showman & Dowling 2000) indicates that they are extremely active dynamically, indicating that the Galileo probe value may not necessarily be representative of the whole planet. In addition, retrievals from infrared spectra indicate that tropospheric temperatures may vary as a function of latitude by up to several kelvins (Fletcher et al. 2020).

Our new retrievals of the Voyager RSS data yield a 1 bar temperature of $170.3 \pm 3.8$ K at 12°S (Voyager 1 ingress) and $167.3 \pm 3.8$ K at 0°N (Voyager 1 egress). Considering the uncertainties, these values are marginally compatible with a uniform temperature equal to the Galileo probe value. However, they also allow for a value that would vary by up to 7 K in this 12°S–7°N equatorial region alone. Note that interior models of Jupiter do better at matching both the Juno and Galileo constraints on heavy-element abundances by invoking higher values of the value of $T_{1bar}$ used as an outer boundary condition, in the range of 175–185 K (Miguel et al. 2022).

Though the Voyager radio occultations cover a wider range of latitudes, longitudes, and times of observation than the Galileo probe, the data for the 1 bar temperature are still limited to the equatorial region. The Voyager ingress and egress latitudes differ by only 12° (Table 1), and their updated temperatures differ by ~3 K at 1 bar (Table 5). Although we cannot be certain that the difference between the revised ingress and egress values is real, they may suggest a warmer-poleward temperature near the equator, which by thermal wind balance would imply the reduction of the strength of the eastward equatorial jet with altitude, consistent with the results of the Galileo Doppler Wind Experiment (Atkinson et al. 1998). Considering Voyager 2, the temperatures at the 1 mb level span a wider range of latitudes. They differ by as much as 20 K (155–175 K) between 0° and 60° S based on Voyager 1 and 2 egress radio occultations done 4 months apart (Tables 8 and 9). Allison (1990) interpreted the large-scale oscillations in the stratospheric temperature retrieved from the Voyager 1 radio occultations as "snapshots" of vertically propagating equatorial waves. The apparent vertical wavelengths of ~3 and 0.5 scale heights can be modeled as the eigenmodes of a leaky waveguide supported by a statically stable region within the deep troposphere, and the corresponding "equivalent depths" are consistent with the phase speeds and latitudinal confinement of the equatorial plumes, hot spots, and other wave features.

Planned radio occultations of Jupiter during Juno's extended mission will cover a wide range of latitudes, including the high latitudes. In the absence of an *S* band on Juno, the temperatures are expected to be retrieved down to the ∼0.5 bar level. The temperature from the updated temperature profiles presented in this paper would be another constraint. Moreover, even at 0.5 bar, the new Juno profiles would provide a means of breaking some of the degeneracies inherent in thermal IR and microwave spectral inversions, strengthening extrapolations to deeper pressures.

The ESA's Jupiter Icy Moons Explorer (JUICE), scheduled for launch in 2023 to operate in the Jupiter system in the 2030s, will carry a radio science instrument (3GM) to be used for downlink occultation experiments (among other things) at the *X* and *Ka* bands. These occultations cover a wide range of latitudes, longitudes, and local times. Combined with the Juno high-latitude occultations, these have the potential to substantially improve our understanding of Jupiter's temperature profile in the years to come.Acknowledgments

We thank Sami W. Asmar, Edwin Bell, Richard G. French, David Hinson, Richard A. Simpson, Linda Spilker, and Edward C. Stone for taking time out to aid us in our search for data and for pointing us in the direction of useful resources. This research was supported by the NASA Juno Project under SWRI Subcontract 699056KC to S.K.A. at the University of Michigan. L.N.F. was supported by a European Research Consolidator Grant (under the European Union's Horizon 2020 research and innovation program, grant agreement No. 723890) at the University of Leicester. Some of this research was carried out at the Jet Propulsion Laboratory, California Institute of Technology, under a contract with the National Aeronautics and Space Administration (80NM0018D0004).

## ORCID iDs

Pranika Gupta https://orcid.org/0000-0002-9566-0372
Sushil K. Atreya https://orcid.org/0000-0002-1972-1815
Paul G. Steffes https://orcid.org/0000-0003-3962-8957
Leigh N. Fletcher https://orcid.org/0000-0001-5834-9588
Tristan Guillot https://orcid.org/0000-0002-7188-8428
Michael D. Allison https://orcid.org/0000-0001-9841-193X
Scott J. Bolton https://orcid.org/0000-0002-9115-0789
Ravit Helled https://orcid.org/0000-0001-5555-2652
Steven Levin https://orcid.org/0000-0003-2242-5459
Cheng Li https://orcid.org/0000-0002-8280-3119
Jonathan I. Lunine https://orcid.org/0000-0003-2279-4131
Yamila Miguel https://orcid.org/0000-0002-0747-8862
Glenn S. Orton https://orcid.org/0000-0001-7871-2823
J. Hunter Waite https://orcid.org/0000-0002-1978-1025
Paul Withers https://orcid.org/0000-0003-3084-4581